\pdfoutput=1
\documentclass[cits]{JINST}

\usepackage{amsmath}
\usepackage{tabularx}
\usepackage{booktabs}

\title{A prototype vector magnetic field monitoring system for a neutron electric
dipole moment experiment}

\author{
N.\ Nouri$^a$\thanks{Corresponding author.}~, A.\ Biswas$^b$, M.\ A.\ Brown$^a$,
R.\ Carr$^b$, B.\ Filippone$^b$, C.\ Osthelder$^b$, B.\ Plaster$^a$, S.\ Slutsky$^b$ and C.\ Swank$^b$\\
\llap{$^a$} Department of Physics and Astronomy, University of Kentucky,\\
Lexington, Kentucky 40506, U.S.A.\\
\llap{$^b$} Kellogg Radiation Laboratory, California Institute of Technology,\\
Pasadena, California 91125, U.S.A.\\
E-mail: \email{nima.nouri@uky.edu}}

\abstract{We present results from a first demonstration of a magnetic
field monitoring system for a neutron electric dipole moment experiment.
The system is designed to reconstruct the vector components of the magnetic
field in the interior measurement region solely from exterior measurements.}


\begin{document}

\section{Introduction}
\label{sec:intro}
In previous work \cite{nouri1,nouri2}, some of us outlined the concept for
a magnetic
field monitoring system for a neutron electric dipole moment experiment in
which the spatial dependence of the vector components of the magnetic field
(and, hence, the gradients $\partial B_i/\partial x_j$) within the inaccessible interior
measurement region are reconstructed solely from exterior measurements.  In 
this paper, we show results from first prototyping tests of such a system.  
The results highlight the potential for the implementation of an improved system 
in an upcoming neutron electric dipole moment experiment to be 
carried out at the Spallation Neutron Source at Oak Ridge National Laboratory
(i.e., the SNS nEDM experiment \cite{ito,Golub}). The work presented here
complements techniques that have been developed to extract interior gradients
from exterior measurements in scalar magnetometers \cite{Afach_1,Afach_2}.

\section{Methodology}
\label{sec:methodology}
Detailed discussions of the concepts underlying our field monitoring system
were published previously \cite{nouri1,nouri2}. In brief, the basic idea 
is as follows. In a region of space containing no sources of 
current or magnetization (key assumptions underlying our concept), 
the magnetic scalar potential obeys the Laplace equation $\nabla^2 \Phi_M(\vec{x})=0$,
for which the solution can be written in spherical coordinates according to
the well-known multipole expansion:	
\begin{equation}\label{eq:general_solution}
\Phi_M\,(r,\theta,\phi) = \sum_{\ell=0}^\infty \sum_{m=0}^{+\ell} r^\ell
P_\ell^m(\cos\theta) \Big[a_{\ell m}\;\cos (m\phi) + b_{\ell m}\;\sin (m\phi) \Big]\,,
\end{equation}
Here, the $P_\ell^m (\cos\theta)$ denote the associated Legendre polynomials,
and $a_{\ell m}$ and $b_{\ell m}$ denote \emph{a priori} unknown expansion coefficients.  
The magnetic field can then, of course, be written as  
$\vec{B} = - \vec{\nabla}\Phi_M(\vec{x})$.
Thus, if values for
the magnetic field components $B_i$ can be determined at a number of exterior
points, such measurements can be employed to determine values for the
expansion coefficients $a_{\ell m}$ and $b_{\ell m}$, which in turn permits 
reconstruction of the scalar potential in the interior region (up to an
arbitrary constant).
Taking the gradient of the expansion equation (\ref{eq:general_solution}) then
uniquely determines the spatial dependence
of the vector field components everywhere within the interior region.

Translating from spherical to rectangular coordinates, we write the components
$B_i(\vec{x})$ in terms of the $a_{\ell m}$ and $b_{\ell m}$
expansion coefficients as
\begin{equation}\label{eq:B_field_component}
B_i\,(\vec{x}) = \sum_{\ell, m}a_{\ell m}\,f_{\ell m,\, i}\,(\vec{x})+
b_{\ell m}\,h_{\ell m,\, i}\,(\vec{x}),
\qquad i \in \{x, y, z\}.
\end{equation}
where we use $f_{\ell m, i}(\vec x)$ and $h_{\ell m, i}(\vec x)$ 
to denote the basis functions in our expansion.
Explicit expressions for $f_{\ell m, i}(\vec x)$ and $h_{\ell m, i}(\vec x)$
written in terms of $(x, y, z)$
rectangular coordinates (i.e., the coordinate system most compatible with
typical experimental field measurements) are given in Appendix B of Ref.\ \cite{nouri2}.

Table \ref{table:basis_fn_table} summarizes the sensitivity of 
$(B_x, B_y, B_z)$ to the $a_{\ell m}$ and $b_{\ell m}$
expansion coefficients (up to $\ell = 3$) in terms of these $f_{\ell m,\, i}$ and $h_{\ell m,\, i}$ 
basis functions.
The point here is that measurements of $(B_x, B_y, B_z)$ provide (mostly)
redundant information on the $a_{\ell m}$ and $b_{\ell m}$ expansion coefficients, 
as most of these coefficients are common to all three 
$(B_x, B_y, B_z)$ components; however,
as can be seen in the table, certain coefficients can only be determined from
measurements of a particular component (e.g., $b_{1 1}$ can only be determined from
a measurement of $B_y$).

If one desires reconstruction of all the $(B_x, B_y, B_z)$ field components
to a certain order (written in terms of the
$a_{\ell m}$ and $b_{\ell m}$ coefficients up to that order), it is possible
via
measurements of any of the $(B_x, B_y, B_z)$ components providing appropriate sensitivity 
to these coefficients, with the number of required measurements equal to
the number of $a_{\ell m}$ and $b_{\ell m}$ coefficients up to that order.
\begin{table}[tb]
    \caption{Sensitivity of the field components $(B_x, B_y, B_z)$ to the 
             $a_{\ell m}$ and $b_{\ell m}$
             expansion coefficients in terms of their associated basis functions,
             $f_{\ell m, i}(\vec x)$ and $h_{\ell m, i}(\vec x)$, in the expansion of the 
             magnetic scalar potential.
             Some trivial basis functions are not listed, e.g., $f_{00,\, i}$, whose 
             associated basis function in the scalar potential expression is a constant 
             and thus does not appear in the expressions for the $B_i$.
             Also, expansion coefficients associated with basis functions which are linearly
             dependent on lower-order terms are redundant and do not need to be determined.
             Note that a `$0$' indicates that particular component offers no
             sensitivity to that particular expansion coefficient.}
  \label{table:basis_fn_table}
\begin{tabularx}{\textwidth}{@{}XXXXXXXXXXXXX} 
      \toprule
      & $\mathbf{a_{10}}$ & $\mathbf{a_{11}}$ & $\mathbf{b_{11}}$
      & $\mathbf{a_{20}}$ & $\mathbf{a_{21}}$ & $\mathbf{b_{21}}$
      & $\mathbf{b_{22}}$ & $\mathbf{a_{30}}$ & $\mathbf{a_{31}}$ 
      & $\mathbf{b_{31}}$ & $\mathbf{b_{32}}$ & $\mathbf{a_{33}}$ \\
         \cline{2-13}\noalign{\smallskip}
        $\mathbf{B_x}$ & $0$           & $-1$          & $0$           & $f_{20,\, x}$ & $f_{21,\, x}$ 
                       & $0$           & $h_{22,\, x}$ & $f_{30,\, x}$ & $f_{31,\, x}$ & $h_{31,\, x}$ 
                       & $h_{32,\,x}$  & $f_{33,\, x}$ \\ [2.0 mm]
        $\mathbf{B_y}$ & $0$           & $0$           & $-1$          & $f_{20,\, y}$ & $0$ 
                       & $h_{21,\, y}$ & $h_{22,\, y}$ & $f_{30,\, y}$ & $f_{31,\, y}$ & $h_{31,\, y}$ 
                       & $h_{32,\, y}$ & $f_{33,\, y}$ \\ [2.0 mm]
        $\mathbf{B_z}$ & $1$           & $0$           & $0$           & $f_{20,\, z}$ & $f_{21,\, z}$ 
                       & $h_{21,\, z}$ & $0$           & $f_{30,\, z}$ & $f_{31,\, z}$ & $h_{31,\, z}$ 
                       & $h_{32,\, z}$ & $0$ \\[2.0 mm]
      \bottomrule
\end{tabularx}
\end{table}
As a specific example, suppose one desired to reconstruct $(B_x, B_y, B_z)$ to order
$(\ell, m) = (3, 3)$ in the scalar potential. As can be inferred from the table,
such a reconstruction would require the determination of only $12$ coefficients,
$a_{1 0}$ through $a_{3 3}$, and, hence, only $12$ exterior measurements. Five of
these coefficients $(a_{20},\ a_{30},\ a_{31},\ b_{31},\ b_{32})$ 
could be determined from a measurement of any of
$(B_x, B_y, B_z)$ at any location in space where their respective basis functions are
non-zero, four of these $(a_{21},\ b_{21},\ b_{22},\ a_{33})$ could be determined 
from any two of $(B_x, B_y, B_z)$,
and three of these $(a_{10},\ a_{11},\ b_{11})$ would require a measurement of a specific component
(e.g., $a_{10}$ is sensitive only to $B_z$). Thus, one can optimize the choice of
external measurements according to the needs of the particular experiment,
with a careful choice of such providing for maximal information in the $(\ell, m)$
reconstruction from a minimal number of exterior measurements.

However, if the experiment is somehow constrained such that exterior
measurements of only one of the field components is possible, then one
can only reconstruct that particular component.  For example, as can be
seen in Table \ref{table:basis_fn_table}, $B_x$ does not provide sensitivity 
to $a_{10}$, $b_{11}$, and $b_{21}$,
coefficients which would be needed for the reconstruction of $B_y$ and $B_z$.
Continuing this example, with $12$ exterior measurements of $B_x$, one could
reconstruct $B_x$ up to a somewhat higher $(\ell, m) = (4, 1)$ order, but would
sacrifice the ability to reconstruct $B_y$ and $B_z$.

\section{Experimental Apparatus and Procedure}
\label{sec:Experimental_apparatus}
A prototype field monitor system consisting of twelve single-axis fluxgate
magnetometer probes was deployed within the magnetic field environment of an
optimized $\cos\theta$ coil surrounded by multiple layers of magnetic
shielding \cite{Perez,Slutsky} developed as a part of prototyping studies for
the SNS nEDM experiment.

\begin{figure}[tb]
\begin{center}
    \includegraphics[width=0.8\textwidth]{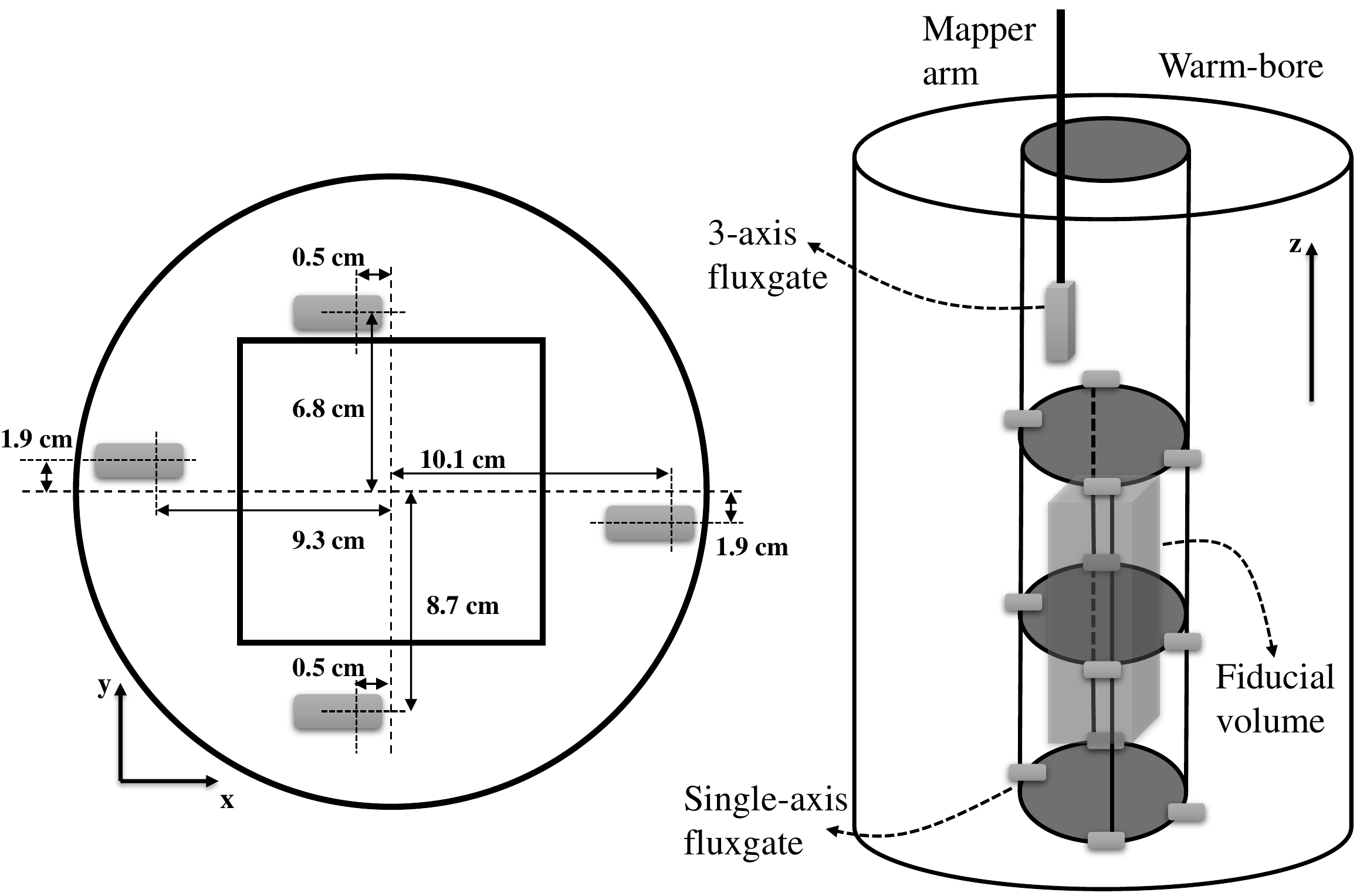}
    \caption{Left panel: A top view of the prototype array design.
             Right panel: Schematic figure of the experimental apparatus. 
             A scaled version of the fiducial volume is also shown.}
  \label{fig:prototype_array}
\end{center}
\end{figure}

The single-axis fluxgate magnetometer probes, obtained from Stefan-Mayer Instruments
\cite{Stefan_Mayer}, were mounted on a cylindrical-like support structure 
consisting of four
aluminum rods, as shown in figure \ref{fig:prototype_array}, with the 
cylindrical axis of this support structure
oriented along the axis of the $\cos\theta$ coil. For the purposes of this
first demonstration, the arrangement of the probes was rather simplistic;
all of the probes were oriented along the $\cos\theta$ coil field direction
(i.e., perpendicular to the axis of the $\cos\theta$ coil), with four probes
mounted (identically) on each of three planes oriented, as shown in 
figure \ref{fig:prototype_array}, perpendicular to the coil axis.

To set the geometric scale for these first tests, note that the radius of
this $\cos\theta$ coil is $32.4$ cm, with a length of $214.6$ cm \cite{Perez}. 
The effective radius
of the support structure for the field monitor probes was $12.85$ cm, with the three
planes separated by a distance of $30.5$ cm.
During data taking, the $\cos\theta$ coil was energized to a nominal field
of $\sim 220$ mGauss at the coil center. The readings in the twelve field
monitor probes were then recorded simultaneously at one point in time.
These twelve readings of the field component along the primary field direction,
which we define to be $B_x$, were then fitted to the multipole expansion in
equation (\ref{eq:B_field_component}) to determine the $a_{\ell m}$ and $b_{\ell m}$
coefficients up to order $(4,1)$, as discussed in the example given in the
previous section. The magnetic
field in the region interior to the prototype array was then calculated using these 
fitted coefficients. The accuracy of the reconstruction was then assessed 
by comparing the results of the reconstruction with explicit measurements
of the field in the interior region carried out using a triple-axis fluxgate
magnetometer probe \cite{Bartington}
mounted to an automated magnetic mapper system.
Note that the readings in the twelve field monitor probes were obtained
prior to the measurements with the automated mapper system; we did not later
average the monitor probes' readings over some period of time.

As is well known, a fluxgate magnetometer probe may read a non-zero value
in a true ``zero field'' environment due to an offset associated with its
geometry and its
electronics. We extracted the offsets of each of our 
single-axis
fluxgate magnetometer probes
(relative to that of the triple-axis probe on the automated mapper system) via a procedure
in which a large number (of order $\sim 700$) of readings from the mapper probe
(conducted with the $\cos\theta$ coil de-energized) obtained in the interior
region of the field monitor probes' support structure were fitted to a
high-order multipole expansion. These fits were then extrapolated to the
single-axis probes' locations, and
any such differences between the extrapolated fitted values and the
single-axis probes' actual readings were
attributed to an offset.

\section{Results}
\label{sec:Results}
\begin{figure}[!tb]
\begin{center}
    \includegraphics[width=\textwidth]{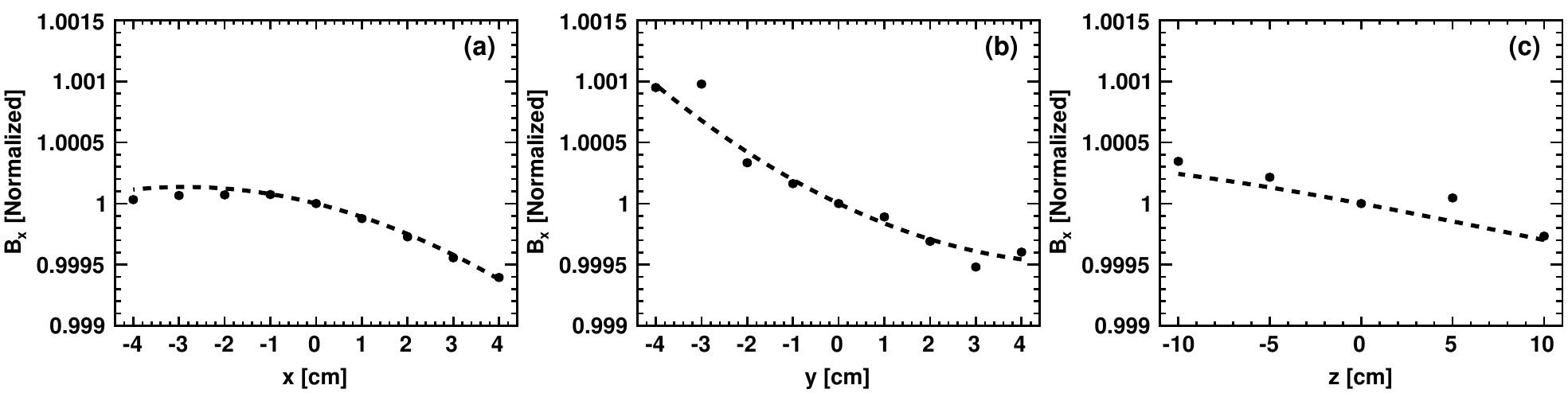}
    \caption{Comparison between the field component $B_x$ (i.e., along primary
    field direction) reconstructed from the exterior measurements in the
    field monitor array (dashed lines) and direct measurements obtained with
    the automated mapper system (dots). The three panels show results along
    the $x$-, $y$-, and $z$-axes.  (Note that the error bars on the data
    points are smaller than the symbol size.)}
  \label{fig:comparison_1}
\end{center}
\end{figure}

Results from our reconstruction of the interior magnetic field
component $B_x$ (i.e., the component oriented along the $\cos\theta$
coil's primary field direction) along the three axes ($z$-axis along
the axis of the $\cos\theta$ coil's cylindrical support structure) are
compared with direct measurements of the field obtained with the field
mapper system in figure \ref{fig:comparison_1}.  Note that the
reconstructed and measured field components $B_x$ were normalized to a
value of $1.0$ at the coil center. As can be seen there, the agreement
is good, with the fractional gradient along the $x$-axis, $(\partial
B_x/\partial x)/B_x$, on the level of $10^{-4}$ cm$^{-1}$.  Note that
whereas the reconstructed fields were based on the readings obtained
in the 12 single-axis probes at some instant in time, the measured
fields (with the single triple-axis probe) were obtained over a time
scale of order $\sim 1$ hour (due to the necessity of having to move
the triple-axis probe to the different locations).  
Thus, the comparison between the reconstructed and measured fields may
be degraded in the presence of time-varying background fields (which
are not completely shielded), either on short time scales (e.g., between
point-to-point measurements of the interior fields by the triple-axis probe),
or on long time scales over the duration of the interior field measurements.
To minimize the possibility of any such issues, the data were taken at night
(data taking during the day was nearly impossible as a result of daytime
operations of an overhead crane and delivery trucks) and the triple-axis probe
was moved point-to-point by the automated mapper system in as expeditious
manner as possible. Nevertheless, as can be seen in figure \ref{fig:comparison_1}, 
some anomalies were observed, such as that at $y = -3$ cm.

Averaged over a volume corresponding to the SNS nEDM experiment's measurement
volume (i.e., over a dense grid of points uniformly filling this volume, of
which a subset are the data points along the central axes shown in figure 
\ref{fig:comparison_1}), the reconstructed and measured fields agreed to
better than $\sim 1.0 \%$. The reconstructed fractional gradients, 
$(\partial B_x/\partial x)/B_x$, $(\partial B_x/\partial y)/B_x$, 
and $(\partial B_x/\partial z)/B_x$, averaged over this half-scale version of
the measurement volume were $-1.1\times 10^{-4}$ cm$^{-1}$, $-1.8\times 10^{-4}$ cm$^{-1}$, 
and $-0.3\times 10^{-4}$ cm$^{-1}$, respectively, which are to be compared to the 
direct measurements, which were $-1.4\times 10^{-4}$ cm$^{-1}$, $-1.2\times 10^{-4}$ cm$^{-1}$, 
and $-0.1\times 10^{-4}$ cm$^{-1}$, again showing good agreement.

Improved results for the SNS nEDM experiment will ultimately be required.
There, the goal will be to monitor the field gradients to the level of 
$10^{-5}$ cm$^{-1}$ or better.  Although this first prototype demonstrating
our method does not yet meet this criterion, we are currently investigating 
the feasibility of employing vector field probes with smaller noise and offset 
characteristics, and in a next-version prototype we will also carry out a full 
optimization of the probe locations for maximal sensitivity to the successive 
higher-order $(\ell, m)$ terms in the multipole expansion.

\acknowledgments
This work was supported in part by the U.\ S.\ Department of Energy Office 
of Nuclear Physics under Award No.\ DE-FG02-08ER41557 and the National 
Science Foundation under Award No.\ 1205977. 
We are grateful to B. E. Allgeier for a careful reading of the manuscript.





\end{document}